\documentstyle{ioplppt}

\def\bfl{\begin{flushleft}}
\def\efl{\end{flushleft}}
\def\bfr{\begin{flushright}}
\def\efr{\end{flushright}}
\def\bc{\begin{center}}
\def\ec{\end{center}}
\def\be{\begin{equation}}
\def\ee{\end{equation}}
\def\ba{\begin{eqnarray}}
\def\ea{\end{eqnarray}}
\def\nn{\nonumber }

\def\text#1{\mbox{#1}}
\def\drm{\text{d}}

\def\schrod{Schr\"oedinger  }

\begin{document}
\jl{1}
\title{{\small 
hep-th/9708018
~~~~~~~~~~~~~~~~~~~~~~~~~~~
J. Phys. A: Math. Gen. 31 (1998) 6081-6085}\\ 
Quantum kink model and SU(2) symmetry:
Spin interpretation and T-violation}[Quantum kink ...]

\author{ K G Zloshchastiev}

\address{ \ Department of Theoretical Physics, 
Dnepropetrovsk State University,
per. Nauchny 13, Dnepropetrovsk 320625, Ukraine}

\begin{abstract}
In this paper we consider 
the class of exact solutions of the \schrod  equation   with 
the Razavi potential.
By means of this we obtain some wavefunctions and 
mass spectra of the relativistic scalar field model with 
spontaneously broken symmetry near the static kink solution.
Appearance of the bosons, which have two different spins, will be shown 
in the theory, thereby the additional breaking of discrete symmetry between 
the quantum mechanical kink particles with the opposite spins
(i.e. the $T$-violation) takes place.
\end{abstract}

At  present, quantum field theories,
having topologically non-trivial solutions are being intensively developed.
Mass spectra  of particles, which are predicted by such theories, 
can be received by means of the effective action formalism 
\cite{coleman}, which describes the low-energy dynamics of stable solutions
taking into account small quantum oscillations.
In particular, \cite{kapust} was devoted to one such 
theory, namely, the $d=1+1$ relativistic model $\varphi^4$ with 
spontaneously broken symmetry.
In that paper the non-perturbative quantum scalar field theory  
near the static kink solution, which can be interpreted as 
a  quantum mechanical heavy particle, was considered.
As a result of quantization of the kink's internal degrees of freedom,
the \schrod   equation was received 
in terms of the raising  and lowering  operators.
It was noted that, dependent on what ordering procedure for the 
operators was chosen, unitary non-equivalent theories take place.
Regrettably, important aspects of the physical sense of such theories, 
as well as the question of obtaining 
exact solutions and mass spectra, remain open.
In this paper we try to resolve these problems in particular.
It became possible owning to the analogies found between the key
equations of \cite{kapust} and the wide class of the \schrod  
equations with the double-well potentials related to SU(2) symmetry  
\cite{brajamani,ghosh,deenen}, in particular, the Razavi potentials   
\cite{razavi,zu,kmr}.

We start from the action
\ba
S[\varphi] = \int \drm^2 x 
\left\{ \frac{1}{2} \frac{\partial \varphi}{\partial x^i}
                    \frac{\partial \varphi}{\partial x_i}
                    -  \frac{1}{4} g^2                     \left[
                          \varphi^2 -                           \left(
                                \frac{m}{g}
                          \right)^2
                    \right]^2
             \right\}       \label{eq1}
\ea
where $\varphi (x,t)$ is the dimensionless scalar field, 
$m$ and $g$ are real parameters.
The corresponding equations of motion have the kink solution \cite{rajaraman}
\ba
\varphi_c(x) = \frac{m}{g} \tanh{
                                 \frac{m x}{\sqrt{2}}
                                } 
\label{eq2}
\ea
with the non-trivial behaviour at infinity
\ba
\varphi_c(+ \infty) = - \varphi_c(- \infty) =\frac{m}{g}  \label{eq3}
\ea
and the non-zero topological charge
\ba
Q =\frac{g}{m} \int\limits^{+\infty}_{-\infty} \frac{\partial \varphi 
(x)}{\partial x} \drm x =\frac{g}{m}
   \left[
         \varphi(x = + \infty) - \varphi(x = - \infty)
   \right].
\ea 
We perform transformation on the new set of the variables
\ba
&&x^m = x^m (s) + e^m_{(1)}(s) \rho  \nn \\ 
&&\varphi(x,t) = \tilde 
\varphi(\sigma_a),~~\sigma_{a=0}=s,~\sigma_{a=1}=\rho 
\label{eq4}
\ea
where $s$ and $\rho$ are the so called collective coordinates, 
$x^m(s)$ turn out to be the coordinates of a (1+1)-dimensional point 
particle, 
$e^m_{(1)}(s)$ is the unit space-like vector, which is orthogonal 
to a worldline of the particle.
It should be pointed out that unlike $(x,t)$ the new base variables 
$(s,\rho)$ are invariant 
under the Poincar\'e transformations.

Considering  field $\tilde \varphi(\sigma_a)$ excitations
near the kink (\ref{eq2}) and 
eliminating  zero modes, it is possible to obtain the
non-minimal $p$-brane (more strictly, non-minimal (1+1)-dimensional
point particle with
curvature) action as a residual effective action 
for the model  (\ref{eq1}), see \cite{kapust} for details,
\ba
&&S_{\text{eff}} = -\mu \int \drm s \sqrt{\dot x^2}             
            \left(
                  1+ \frac{k^2}{3 m^2}
            \right)                                            \label{eq5} \\
&&\mu = \frac{2 \sqrt{2}}{3} \frac{m^3}{g^2}                    \label{eq6}
\ea
where $k=\sqrt{-a^i a_i}$ is the curvature of a point-particle worldline, 
$a_n$ is the acceleration
\[
a_n=\frac{1}{\sqrt{\dot x^2}}
    \frac{\drm}{\drm s}
    \frac{\dot x_n}{\sqrt{\dot x^2}}.                            
\]

From (\ref{eq5}) it follows that we have obtained a theory with higher 
derivatives.
In this theory we have two pairs of canonical variables 
$\{ x_m,p_m \}$ and $\{ q_m=\dot x_m,\Pi_m \}$ which are constrained 
on a certain submanifold of the total phase space
by both the two primary first-kind constraints  $\Phi_{1,2} \approx 0$ 
and the proper time gauge condition $\sqrt{q_m q^m} \approx 1$.
After some transformations one of the constraints can be rewritten as 
\ba
\Phi_2 &=& - \sqrt{p^2} \cosh{v} + \mu -
             \frac{\mu }{\xi^2} \Pi_v^2 \approx 0              \label{eq7} \\
\xi &=& \frac{2}{ \sqrt{3}}
        \frac{\mu}{m}                                           \label{eq8}
\ea
where $v$ is the new coordinate, $\Pi_v$ is the corresponding momentum, 
which are interpreted in paper \cite{kapust} as the spin values.
Below it will be shown that this interpretation is 
not complete and the true SU(2) spin operators will be introduced.

In the quantum case  the condition $~\widehat \Phi_2 |\Psi \rangle =0$,
where  $~\widehat \Pi_v = -i~ \partial / \partial v$ is the momentum 
operator in the coordinate representation and 
$\Psi=\Psi (v)$ is the wavefunction of the kink, must be satisfied.
The constraint $\widehat \Phi_2$ permits two modifications,
consideration of which gives us the equations of 
motion in terms of the raising  and lowering  operators
\ba
[ a^{\dagger}_{\lambda} a^{~}_\lambda - \xi^2 (1-\lambda) ]
                                             \Psi (v) = 0      \label{eq9} \\
~[ a^{~}_{\lambda} a^{\dagger}_\lambda - \xi^2 (1-\lambda) ]
                                             \Psi (v) = 0       \label{eq10}
\ea
where
\ba
&&a_{\lambda} = \frac{\drm}{\drm v} + \sqrt{2\lambda} \xi 
       \sinh{ \frac{v}{2}}                                    \label{eq11} \\
&&\lambda = \frac{\sqrt{p^2}}{\mu}  = \frac{M}{\mu}.          \label{eq12}
\ea

These equations can be written in the form of the \schrod  equation 
\ba
&&\left(
  -\frac{\drm^2}{\drm v^2} +
  2\lambda \xi^2
  \sinh{\! ^2\frac{v}{2}}
  -\sqrt{\frac{\lambda}{2}}~ \xi
  \cosh{\frac{v}{2}}
\right) \Psi (v) = \xi^2 (1-\lambda) \Psi (v)               \label{eq13}\\
&&\left(
  -\frac{\drm^2}{\drm v^2} +
  2\lambda \xi^2
  \sinh{\! ^2\frac{v}{2}}
  +\sqrt{\frac{\lambda}{2}}~ \xi
  \cosh{\frac{v}{2}}
\right) \Psi (v) = \xi^2 (1-\lambda) \Psi (v). \label{eq14}
\ea

In \cite{kapust} only equation (\ref{eq13}) was considered, 
for which the
wavefunction of the ground (vacuum) state was found  
\ba
&&a_{\lambda=1} \Psi_{\text{vac}}(v) = 0  \nn \\ 
&&\Psi_{\text{vac}}(v) = C \exp{
                        \left(
                          -2 \sqrt{2} \xi
                          \cosh{
                                 \frac{v}{2}
                               }
                        \right)
                       }.                                        \label{eq15}
\ea 

Below we represent the approach, which  helps us to take a new look at the 
expressions (\ref{eq9}) - (\ref{eq14}) as well as to obtain some exact 
results and deeper interpretation of the theory.

Let us consider the \schrod  equation   \cite{razavi,zu,kmr}
\ba
[ \widehat H-\varepsilon] \Psi(\zeta) = 0  \label{eq16} 
\ea 
where
\ba
\widehat H =  -\frac{\drm^2}{\drm \zeta^2} +
  \frac{B^2}{4}
  \sinh{\! ^2 \zeta}
  -B   \left(
        S+\frac{1}{2}
  \right)
  \cosh{\zeta}.                                             \label{eq17}
\ea

Here $S$ and $B$ are dimensionless parameters.
It can readily be shown that SU(2) is the dynamic group of symmetry for this 
Hamiltonian and to provide the direct analogy with the spin Hamiltonian
\ba
\widehat H_s= -S^2_z - B S_x                                 \label{eq18}
\ea
using the information induced by the su(2) Lie algebra \cite{tur}.

On a subset $L^2(R)$  the following spin operators act
\ba
&&S_x = S \cosh{\zeta} - \frac{B}{2} \sinh{\!^2 \zeta} - \sinh{\zeta} 
\frac{\drm}{\drm\zeta}  \label{eq19}\\
&&S_y = i         \left\{
               -S \sinh{\zeta} + \frac{B}{2} \sinh{\zeta}\cosh{\zeta} + 
\cosh{\zeta} \frac{\drm}{\drm\zeta} \right\}  \label{eq20}\\
&&S_z =          \frac{B}{2} \sinh{\zeta}
        + \frac{\drm}{\drm\zeta}.                                      \label{eq21}
\ea

Thereby the commutation relations
\ba
&&[S_i,~S_j] = i \epsilon_{ijk} S_k                            \label{eq22}\\
&&S_x^2+S_y^2+S_z^2 \equiv S (S+1)                             \label{eq23}
\ea
are valid.

Now we consider the case $S\geq 0$.
Then an irreducible finite-dimensional subspace of representation 
space of the SU(2) algebra, which is
invariant with respect to the operators (\ref{eq19})-(\ref{eq21}), exists.
Its dimension is $2S+1$.

One can verify \cite{zu} that the solution of (\ref{eq16}) is the 
function
\ba
\Psi (\zeta) =\exp{
                   \left(
                         -\frac{B}{2} \cosh{\zeta}
                   \right)
                  }
              \sum_{\sigma=-S}^{S}
              \frac{c_\sigma}
                   {
                    \sqrt{
                          (S-\sigma)\verb|!|~
                          (S+\sigma)\verb|!|
                         }
                   }
              \exp{
                   \left(
                         \sigma \zeta
                   \right)
                  }                                              \label{eq24}
\ea
where the coefficients $c_\sigma$ satisfy with the system of linear 
equations
\ba
&&\biggl(
       \varepsilon+\sigma^2
\biggr)c_\sigma + \frac{B}{2}
\biggl[
       \sqrt{(S-\sigma)(S+\sigma+1)}~ c_{\sigma+1}              \nn \\
&&+ \sqrt{(S+\sigma)(S-
\sigma+1)}~ c_{\sigma-1}
\biggr] = 0                                                 \label{eq25} \\
&&c_{S+1} = c_{-S-1}=0~~~\sigma=-S,~-S+1,...,~S.             \nn
\ea

The solution of this system is equivalent to the determination of 
eigenvectors
and eigenvalues of the operator $\widehat H$ in the matrix representation,
which is realized in a finite-dimensional subspace of  
the su(2) Lie algebra.
Analytical solutions of equations
(\ref{eq24}) and (\ref{eq25}) were found for the following spin values.

(i) $S=0$. 
The dimension of the invariant subspace of the algebra is equal to 1, 
therefore, only one wavefunction and 
ground-state energy can be found.
We have
\ba
\Psi_0 (\zeta) = A_0     
        \exp{
              \left(
                    -\frac{B}{2} \cosh\zeta
              \right)
             },~\varepsilon_0=0.                               \label{eq26}
\ea

(ii) $S=\frac{1}{2}$. 
We obtain two wavefunctions and energies of according states
\ba
&&\Psi_0 (\zeta)
     = A_0
         \exp{
              \left(
                    -\frac{B}{2} \cosh\zeta
              \right)
             }
         \cosh{
               \left(
                     \frac{1}{2} \zeta
               \right)
              }~~~
         \varepsilon_0=
         -\frac{B}{2}
         -\frac{1}{4}                                                  \nn \\
&&\Psi_1 (\zeta)
     = A_1
         \exp{
              \left(
                    -\frac{B}{2} \cosh\zeta
              \right)
             }
         \sinh{
               \left(
                     \frac{1}{2} \zeta
               \right)
              }~~~
         \varepsilon_1=
         \frac{B}{2} -\frac{1}{4}.
\ea

(iii) $S=1$.
There are solutions for three lower levels
\ba
&&\Psi_0 (\zeta)  = A_0          \exp{               
              \left(
                    -\frac{B}{2} \cosh\zeta
              \right)
             }
         \left(
                1-\frac{\varepsilon_0}{B}\cosh{\zeta}
         \right)~~~
         \varepsilon_0=
         -\frac{r_+}{2}                                                \nn \\
&&\Psi_1 (\zeta)   = A_1          
         \exp{               
              \left(
                    -\frac{B}{2} \cosh\zeta
              \right)
             }
         \sinh{\zeta}~~~
         \varepsilon_1=-1                                     \label{eq27}\\
&&\Psi_2 (\zeta) = A_2          
         \exp{               
              \left(
                    -\frac{B}{2} \cosh\zeta
              \right)
             }
         \left(
                1-\frac{\varepsilon_2}{B}\cosh{\zeta}
         \right)~~~
         \varepsilon_2=
         -\frac{r_-}{2}                                         \nn 
\ea 
where $r_\pm=1 \pm \sqrt{1+4 B^2}$, $A_i$ are integration constants.

(iv) $S = \frac{3}{2},~2$.
In this work these cases are not considered, however, it is still
possible to find exact solutions for them  \cite{zu}.

(v) $S > 2,~~2S$ is an integer.
Since it is impossible to solve the system (\ref{eq25}) exactly,
there are not any analytical solutions in this case.

(vi) Either $2S$ is a non-integer or $S<0$.
For such spin values an invariant subspace of the algebra does not 
exist \cite{kmr}.

Now we apply the results obtained above to the $d=1+1$ 
relativistic model $\varphi^4$ with spontaneously
broken symmetry near the static kink solution.
Indeed, it is easy to show that the equations (\ref{eq13}) and 
(\ref{eq14}) can be 
rewritten in the form (\ref{eq16}), if we suppose 
\ba 
v=2 \zeta~~~B=4\sqrt{2 \lambda}\xi~~~\varepsilon=4\xi^2(1-\lambda)  \nn
\ea
and $S=0$ for the equation (\ref{eq13}), $S=-1$ for (\ref{eq14}).

For $S=0$ from equations (\ref{eq8}), (\ref{eq12}) and (\ref{eq26}) we 
obtain the mass  spectrum of the kink boson in the ground state.
It equals to the spectrum for the case of a free particle with the 
mass $\mu$
\ba
M^{(S=0)}_{(n=0)} = \mu 
\ea
and the known result (\ref{eq15}) can be obtained.

It is easy to see that the symmetry between the
particles with the spins $S=1$ and $S=-1$ is broken in the studied theory.
Indeed, for $S=1$  
the equations (\ref{eq8}), (\ref{eq12}) and (\ref{eq27}) yield
\ba 
&&M^{(S=1)}_{(n=0)} =\mu -
                    \frac{21}{32} \frac{m^2}{\mu}
                    \pm
                    \frac{27 m}{32}                     
                    \sqrt{
                           \left(
                                 \frac{m}{\mu} 
                           \right)^2 + \frac{512}{243} 
                          }                                       \nn \\ 
&&M^{(S=1)}_{(n=1)} = \mu + \frac{3}{16} \frac{m^2}{\mu}  \\ 
&&M^{(S=1)}_{(n=2)} = \mu +
                    \frac{27}{32} \frac{m^2}{\mu}
                    \pm
                    \frac{27 m}{32}                     
                    \sqrt{
                           \left(
                                 \frac{m}{\mu} 
                           \right)^2
                           + \frac{512}{243}
                         }                                        \nn    
\ea
where the signs `$\pm$' denote additional 
splitting of even mass spectra  at least for  lower states.
Therefore, in this theory the breaking of discrete symmetry under the 
time inversion (the $T$-violance) takes place \cite{ryder} additionally
to the symmetry breaking of the initial theory (\ref{eq1}). 

It should also be noted  that the true spin operator of the model is not 
the operator of the canonical momentum $\widehat\Pi_v$. 
The true spin operators are given by the 
expressions (\ref{eq19}) - (\ref{eq21}),
thereby for the case $S=0$ they are equal to $S_z$ up to
the factors $-\sinh{\zeta}$ and $i \cosh{\zeta}$ respectively.

Finally, we represent another important aspect.
It is now evident that the main demand on
models such as that of \cite{kapust} is their correspondence 
to reality.
Otherwise all these particle-like solutions and field theories  
based on them  will be no more than interesting mathematical toys, 
i.e. `physics for one day'.
As for the model considered here the analogy  with 
the spin Hamiltonians is very useful in this connection.
It is well known that both the Hamiltonians like (\ref{eq18}) and double-well 
potentials are often exploited in physics.
As examples one can point out the following applications, 
the anisotropic paramagnet \cite{zuts}, 
the theory of molecular vibrations \cite{volkenshtein}, 
the model of the anharmonic oscillator in field theories \cite{bender}, 
and 
the model of interacting fermions  in nuclear physics \cite{lipkin}.

\def\JETPr{Zh. Eksp. Teor. Fiz.}

\end{document}